\documentclass[sigconf]{acmart} 

\copyrightyear{2026}
\acmYear{2026}
\setcopyright{rightsretained}
\acmConference[ICSE '26]{2026 IEEE/ACM 48th International Conference on Software Engineering}{April 12--18, 2026}{Rio de Janeiro, Brazil}
\acmBooktitle{2026 IEEE/ACM 48th International Conference on Software Engineering (ICSE '26), April 12--18, 2026, Rio de Janeiro, Brazil}
\acmDOI{10.1145/3744916.3764575}
\acmISBN{979-8-4007-2025-3/26/04}

\usepackage{multirow}
\usepackage{listings}
\usepackage[ruled,vlined,linesnumbered,lined]{algorithm2e}
\usepackage[shortlabels]{enumitem}

\newcommand{\tool}{\textsc{Synedrion}}

\newsavebox{\answerboxbox}
\newenvironment{answerbox}{%
  \begin{lrbox}{\answerboxbox}%
    \begin{minipage}{0.97\linewidth}%
    \ignorespaces
}{%
    \end{minipage}%
  \end{lrbox}%
  \noindent\fcolorbox{gray!50}{gray!5}{\usebox{\answerboxbox}}%
}

\newsavebox{\grayframebox}

\author{Sina Gogani-Khiabani}
\orcid{0000-0002-1060-1753}
\affiliation{%
  \institution{University of Illinois Chicago}
  \city{Chicago, IL}
  \country{USA}
}
\email{sgoga3@uic.edu}

\author{Ashutosh Trivedi}
\orcid{0000-0001-9346-0126}
\affiliation{%
  \institution{University of Colorado Boulder}
  \city{Boulder, CO}
  \country{USA}
}
\email{ashutosh.trivedi@colorado.edu}

\author{Diptikalyan Saha}
\orcid{0000-0002-1583-5479}
\affiliation{%
  \institution{IBM Research}
  \city{Bangalore}
  \country{India}
}
\email{diptsaha@in.ibm.com}

\author{Saeid Tizpaz-Niari}
\orcid{0000-0002-1375-3154}
\affiliation{%
  \institution{University of Illinois Chicago}
  \city{Chicago, IL}
  \country{USA}
}
\email{saeid@uic.edu}

\begin{document}

\title{An LLM Agentic Approach for Legal-Critical Software: A Case Study for Tax Prep Software}

\begin{abstract}
As Large Language Models (LLMs) continue to advance in their ability to process both natural and programming languages, they show promise for translation tasks in domains with strict compliance requirements. Yet ensuring consistency in legally critical settings remains challenging due to inherent limitations such as natural language ambiguity and the tendency to hallucinate. This paper explores an \emph{agentic approach} that leverages LLMs for the development of legal-critical software. We use \emph{U.S. federal tax software} as a representative case study, where natural language tax code must be translated precisely into executable logic.

A central challenge in developing legal-critical software from specifications lies in test case generation, which suffers from the \emph{oracle problem}: determining the correct output for a given scenario often requires interpreting legal statutes. 
Prior work has proposed \emph{metamorphic testing} as a solution by evaluating equivalence across similarly situated individuals. A key innovation of our work is a \emph{higher-order} generalization of \emph{metamorphic tests}, motivated by our tax preparation case study, in which system outputs are compared across structured shifts among similar individuals. 
Since manually generating such higher-order relations is tedious and error-prone, our agentic paradigm is well suited to automate test case generation.

We design and implement \tool{}, an assembly of LLM-based agents simulating roles in real-world software development teams handling legal documents. The framework includes a \emph{metamorphic testing agent} that produces counterexamples while translating tax code into executable software. Our findings indicate that \tool{}, when employing smaller language models (e.g., \texttt{GPT-4o-mini}), can outperform frontier models (e.g., \texttt{GPT-4o} and \texttt{Claude-3.5}) in complex tax code generation tasks, achieving a worst-case pass rate of 45\% compared with 9\%--15\%. We thus make the case for LLM-driven agentic methodologies as a pathway for generating robust, trustworthy legal-critical software from natural language specifications.
\end{abstract}

\ccsdesc[500]{Software and its engineering~Software verification and validation}
\ccsdesc[300]{Software and its engineering~Automated static analysis}
\ccsdesc[500]{Computing methodologies~Natural language processing}
\ccsdesc[300]{Information systems~Decision support systems}
\ccsdesc[300]{General and reference~Law}

\keywords{LLMs, Agentic AI, Tax Preparation Software, Metamorphic Testing}

\maketitle

\section{Introduction}
\label{sec:intro}
Large Language Models (LLMs) have the potential to transform legal-critical software development by automating the interpretation and implementation of complex regulatory requirements. 
Legal-critical systems in domains such as finance, healthcare, and compliance demand highly accurate translations of natural language statutes into executable code. 
Ensuring correctness, consistency, and compliance in these domains is particularly challenging due to the intricacy of legal language, the frequency of regulatory updates, and the absence of explicit oracles for validation. 
Meeting these challenges requires principled methodologies that both minimize errors and fully exploit the capabilities of LLMs. 
In this work, we introduce an \emph{agentic approach} to LLM-driven legal-critical software development, using U.S. federal tax preparation software (\emph{tax software}) as a representative case study.

\subsubsection*{Tax Preparation Software.}  
With an estimated 72 million Americans relying on tax software to file their returns in 2020, tax preparation systems exemplify legal-critical software due to their stringent compliance requirements and central role in enforcing tax laws~\cite{market-size,IRS-efile,ICSE-SEIS23,goganikhiabani2025technicalchallengesmaintainingtax}. Ensuring compliance while maintaining usability makes the development of such systems inherently complex~\cite{10.1145/3306618.3314279,EscherB20,ICSE-SEIS23,gogani2025performance}, requiring expertise in both mission-critical software engineering and legal interpretation.  
Tax software must faithfully implement and update the tax code---a structured body of statutes and regulations governing the computation of taxes for individuals and businesses. Translating these provisions into executable logic poses multiple challenges, including statutory ambiguity, intricate eligibility criteria, and interdependencies among tax rules. The evolving nature of tax law further demands frequent software updates, which heightens the risk of implementation errors.  
Such errors are well documented: for example, OpenTaxSolver misapplied Earned Income Tax Credit (EITC) eligibility rules~\cite{ICSE-SEIS23}, and TaxSlayer blocked eligible taxpayers from claiming Affordable Care Act (ACA) credits~\cite{ACA-Error}, underscoring the difficulty of faithfully encoding legal regulations into compliant software.  
Given these complexities---from legal ambiguity to continual regulatory change---tax software provides an ideal testbed for exploring LLM-based legal software development, where LLMs can support the interpretation, updating, and compliance with evolving tax law.

\subsubsection*{Metamorphic Testing.}  
A fundamental principle of common law, {\it stare decisis}, dictates that similar cases should yield similar rulings. 
This ensures consistency and fairness by binding decisions to precedent and comparative reasoning. 
It also enables the expression of legal properties through case comparisons---an approach that naturally aligns with metamorphic testing in software validation. 
Following Tizpaz-Niari et al.~\cite{ICSE-SEIS23}, we connect {\it stare decisis} to metamorphic testing as a means of validating legal-critical software.  

Metamorphic testing addresses the oracle problem---where the correct output is unknown---by ensuring relative correctness between input pairs. 
This mirrors {\it stare decisis} by enforcing consistency across similar cases. 
For instance, while a taxpayer’s exact liability may be uncertain, we can assert that a blind taxpayer should receive a higher deduction than an otherwise identical non-blind taxpayer. 
Formally: \emph{For any two individuals identical except for blindness, the blind taxpayer’s liability must be less than or equal to that of the non-blind taxpayer.}  

Prior research~\cite{ICSE-SEIS23} has applied metamorphic testing to tax software by comparing outputs across similarly situated individuals rather than relying on absolute correctness. 
However, conventional metamorphic testing, which relies on pairwise comparisons, may overlook systematic errors. 
A case in point is \emph{progressive taxation}: a faulty tax program that applies a flat tax rate across all incomes may still satisfy pairwise metamorphic relations (e.g., higher income leading to higher tax) while violating the principle of progressivity.
 
To address these limitations, we introduce \emph{higher-order metamorphic relations} evaluating rates of change across multiple taxpayer profiles. 
Instead of verifying that a higher-income taxpayer owes more tax than a lower-income one, our method examines whether the rate of tax increase reflects statutory progressive structures.

Since manually specifying higher-order metamorphic relations is tedious and error-prone, we leverage LLM agents trained to infer metamorphic properties from legal documents, structured examples, and domain-specific heuristics. 
Metamorphic testing thus provides a structured way to enforce legal consistency by grounding validation in comparative properties rather than absolute correctness. 
Our higher-order metamorphic testing agent extends this framework by capturing broader legal reasoning patterns and detecting systematic discrepancies that traditional pairwise comparisons may miss.

\begin{figure}[t]
    \centering
    \includegraphics[width=0.99\linewidth]{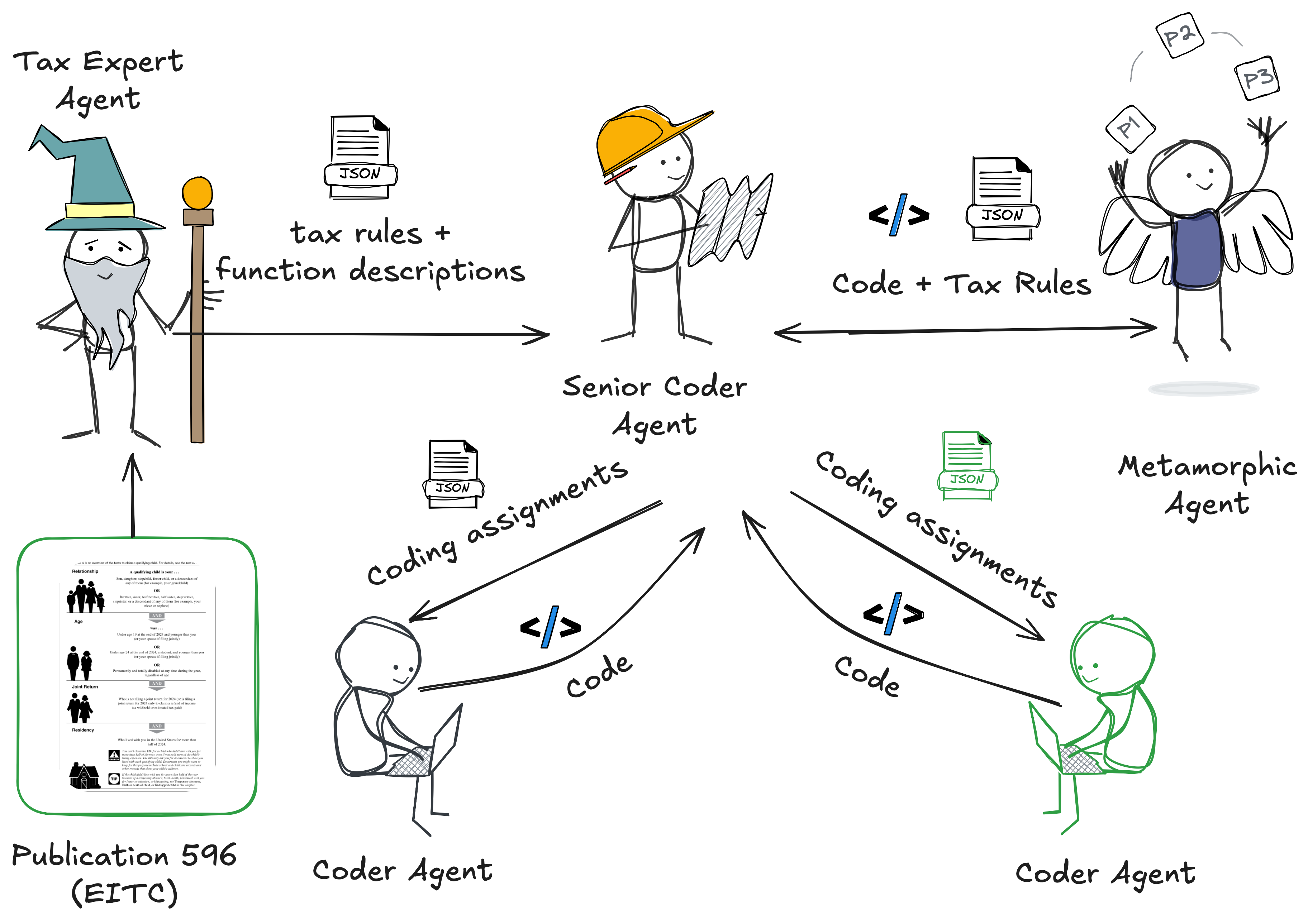}
    \Description{Diagram of the proposed LLM multi-agent framework \tool{} showing modules
    for parsing legal documents, coordination among agents, and generation of executable tax software.}
    \caption{\tool{}---an LLM multi-agent framework for implementing tax software from legal documents.}
    \label{fig:framework}
\end{figure}

\subsubsection*{LLMs for Code Generation.}
LLMs---such as ChatGPT, Llama, and Claude---have demonstrated rapid advancements in capabilities and applications. Trained on vast corpora of text, including books, web pages, and publicly available code, these models are increasingly applied to software engineering tasks~\cite{hindle2016naturalness,fan2023large,li2023starcoder}. However, LLMs face significant challenges, including token constraints, susceptibility to hallucinations, and the absence of a persistent state across sessions. 
More critically, LLMs struggle with the inherent ambiguity of software requirements expressed in natural language.
For instance, our experiments reveal that when prompted to $\mathtt{``calculate\ tax\ based\ on\ progressive\ brackets"}$, an LLM may incorrectly apply the highest tax rate to an entire income rather than only to the portion exceeding lower bracket thresholds. Such errors highlight a fundamental disconnect between natural language specification, code generation, and verification.

\subsubsection*{Our Approach.}  
We posit that the structured, formal nature of legal language in tax amendments, combined with concrete examples, makes tax-related content particularly suitable for automatic translation into executable software via LLMs. 
Recent work~\cite{liu2024large} has highlighted the potential of LLMs to assume traditional roles in software development, motivating an agentic approach~\cite{Bandura2001,liu2024large} in which specialized agents---each with defined intentions, responsibilities, and actions---collaborate in system design.  

We introduce \tool{}, a \emph{multi-agent framework} (Figure~\ref{fig:framework}) for translating legal documents into executable software. 
A central innovation is the \emph{higher-order metamorphic testing agent} (HMT), which improves robustness by detecting systematic errors in LLM-generated tax software. 
We further provide an \emph{empirical analysis} of six legal tax publications, assessing the contributions of individual agents within our framework and evaluating their effectiveness across different LLM capabilities and tax code complexities.

\subsubsection*{Results.}  
Our study shows that baseline LLMs struggle to generate accurate tax software directly from legal code. 
In contrast, our multi-agent framework demonstrates that smaller models can match or outperform frontier baselines. 
Addressing RQ1--RQ4, we find that while strong single models perform well in simpler cases, they falter under complexity; our multi-agent design sustains performance across scenarios. 
Higher-order metamorphic testing further strengthens robustness by uncovering systematic errors conventional testing overlooks. 
Finally, empirical analysis highlights the role of domain-specific agents, such as \emph{TaxExpertAgent} and \emph{MetamorphicTestingAgent}, in achieving substantial gains in functional correctness. 
Together, these results underscore the benefits of structured agent collaboration for legally critical software.

\section{Tax Software and Metamorphic Relations}
\label{sec:problem}

For convenience, we abstractly represent a functional model of tax software as a tuple \( (X, \mathcal{F}) \), where:
\begin{itemize}
    \item \( X = \{ X_1, X_2, \ldots, X_n \} \) is the set of variables corresponding to fields on an individual's tax return form. 
    These variables include \(\texttt{age}\), a numerical variable for the individual's age; \(\texttt{blind}\), a Boolean variable indicating whether the individual is blind; and \(\texttt{sts}\), the filing status with values such as MFJ (married filing jointly) and MFS (married filing separately).
    \item \( \mathcal{F}: \mathcal{D}_1 \times \mathcal{D}_2 \times \cdots \times \mathcal{D}_n \rightarrow \mathbb{R} \) is the function computed by the software, where each \( \mathcal{D}_i \) is the domain of variable \( X_i \). We write \( \mathcal{D} \) for the Cartesian product \( \mathcal{D}_1 \times \mathcal{D}_2 \times \cdots \times \mathcal{D}_n \).
\end{itemize}

\noindent For an individual taxpayer \( x \in \mathcal{D} \), we write \( x(i) \) for the value of the \( i \)-th variable, or \( x.\texttt{lab} \) for the value of the variable labeled \texttt{lab}. 
Since the ground-truth outcome for any input \( x \) is difficult to predict exactly, testing the functional correctness of tax software encounters the \emph{oracle problem}~\cite{6963470}, wherein the expected output for a given input is unavailable.  

To address this, we draw on metamorphic testing for tax software~\cite{ICSE-SEIS23,tizpazniari2024metamorphicdebugging,padhye2024software}. 
The key observation is that while exact output for a given taxpayer profile may be unknown, the relationship between outputs for related taxpayers can be expressed. 
For example, rather than requiring an oracle for a blind taxpayer, one can compare two individuals differing only in blindness, and require the software to assign the blind individual a lower tax burden.

Let \( \mathcal{L} \) be the set of all labels.  
Given \( L \subseteq \mathcal{L} \) and an individual \( \mathbf{x} \in \mathcal{D} \), we often wish to express properties relating to a counterfactual individual \(\mathbf{x'} \in \mathcal{D}\) who is identical to \( \mathbf{x} \) except possibly on the labels in \( L \).  
We denote this relationship as \( \mathbf{x} \equiv_{L} \mathbf{x'} \), meaning that \(\mathbf{x}\) and \(\mathbf{x'}\) are similar with respect to all labels outside \( L \). Formally,
\[
\mathbf{x} \equiv_{L} \mathbf{x'} \;\;\implies\;\; \forall \ell \notin L, \; \mathbf{x}.\ell = \mathbf{x'}.\ell.
\]
For instance, consider the higher standard deduction available to blind individuals.  
We can express the corresponding relation as:
\[
\forall \mathbf{x}, \mathbf{x'} \;\; \Big( \mathbf{x} \equiv_{\{\texttt{blind}\}} \mathbf{x'} \;\wedge\; \mathbf{x}.\texttt{blind} \Big) 
\;\;\implies\;\; \mathcal{F}(\mathbf{x}) < \mathcal{F}(\mathbf{x'}).
\]

\subsubsection*{Higher-Order Metamorphic Relations.}  
In the context of tax software, prior work~\cite{ICSE-SEIS23,tizpazniari2024metamorphicdebugging} has primarily relied on pairwise metamorphic relations, where two inputs are compared based on an expected directional change. 
While this approach is sound---any violation directly signals a bug---it may fail to capture more complex discrepancies spanning multiple test cases.  

To increase coverage, we extend beyond quaternary relations (relating two inputs and outputs) and adopt \emph{n-ary metamorphic relations}~\cite{ahlgren2021testing}. 
These are expressed as
\[
\phi(\mathbf{x}_1, \mathbf{y}_1, \ldots, \mathbf{x}_n, \mathbf{y}_n),
\]
where \((\mathbf{x}_1,\ldots,\mathbf{x}_n)\) are base test cases, \((\mathbf{y}_1,\ldots,\mathbf{y}_n)\) are follow-ups, and each pair \((\mathbf{x}_i, \mathbf{y}_i)\) satisfies an equivalence constraint \(\mathbf{x}_i \equiv_{L} \mathbf{y}_i\) for some subset of labels \(L \subseteq \mathcal{L}\).  
This higher-order formulation broadens metamorphic testing by evaluating functional requirements across multiple related cases simultaneously, enabling detection of systematic errors that pairwise comparisons might overlook.

\subsubsection*{Motivating Example.}  
To illustrate higher-order metamorphic relations with continuous numerical inputs, consider the U.S. progressive tax system as outlined in IRS Publication~17. 
Tax liability is not a linear function of income but follows a bracketed structure, where portions of income are taxed at different rates. 
For example, a \emph{single-status taxpayer} is taxed at \textbf{10\%} on income up to \textbf{\$11,600} (\( b_1 \)) and at \textbf{12\%} on income from \textbf{\$11,601} to \textbf{\$47,150} (\( b_2 \)). 
Thus, an individual earning \textbf{\$20,000} owes:
\begin{itemize}
    \item \textbf{\$1,160} on the first \textbf{\$11,600} (\(10\%\)),
    \item \textbf{\$1,008} on the remaining \textbf{\$8,400} (\(12\%\)),
    \item \textbf{Total: \$2,168}.
\end{itemize}
A \emph{4-ary relation} can capture a monotonicity property:
\[
\phi_4: \forall \mathbf{x}, \mathbf{x'} \quad (\mathbf{x} \equiv_{I} \mathbf{x'}) \wedge (\mathbf{x}.I \geq \mathbf{x'}.I) \implies \mathcal{F}(\mathbf{x}) \geq \mathcal{F}(\mathbf{x'}).
\]
This ensures that higher income implies at least as much tax. 
However, it \emph{fails to detect} systematic errors. 
For instance, a faulty implementation applying a flat \textbf{12\%} tax to all income (e.g., taxing \textbf{\$20,000} at \textbf{12\%} = \textbf{\$2,400} instead of \textbf{\$2,168}) would still satisfy monotonicity. 
Likewise, for deductions (e.g., medical expenses), pairwise checks can confirm that liability decreases as expenses increase, but not whether the decrease follows the correct statutory proportion.
To capture such errors, we introduce an \emph{8-ary relation} that compares rates of change across multiple income levels:
\begin{multline*}
\phi_8: \forall \mathbf{x_1}, \mathbf{x_2}, \mathbf{y_1}, \mathbf{y_2}, \; 
(\mathbf{x_1} \equiv_{I} \mathbf{y_1}) \wedge (\mathbf{x_2} \equiv_{I} \mathbf{y_2}) \wedge \\
(\mathbf{x_1}.I = \mathbf{x_2}.I < \mathbf{y_1}.I < \mathbf{y_2}.I \in [\$11{,}601{-}\$47{,}150]) \\
\implies \Big| \tfrac{\mathcal{F}(\mathbf{x_1})-\mathcal{F}(\mathbf{y_1})}{\mathbf{x_1}.I-\mathbf{y_1}.I} - \tfrac{\mathcal{F}(\mathbf{x_2})-\mathcal{F}(\mathbf{y_2})}{\mathbf{x_2}.I-\mathbf{y_2}.I} \Big| < 0.12.
\end{multline*}

This ensures that the marginal tax rate remains within the expected bracket, rejecting flat-rate implementations. 
A slight variation further enforces \emph{within-bracket consistency}, requiring taxpayers in the same bracket to face the same rate:
\[
\frac{\mathcal{F}(\mathbf{x_1}) - \mathcal{F}(\mathbf{y_1})}{\mathbf{x_1}.I - \mathbf{y_1}.I} \approx \frac{\mathcal{F}(\mathbf{x_2}) - \mathcal{F}(\mathbf{y_2})}{\mathbf{x_2}.I - \mathbf{y_2}.I}.
\]
Such higher-order relations provide a stronger criterion, catching systematic errors that simpler monotonicity checks cannot.

\section{Tax Software from Legal Documents}
\label{sec:overview}

Given a tax software system \(\mathcal{F}\) synthesized from the natural language of the tax code, our work addresses the fundamental challenge of verifying its functional correctness against legal requirements. 
The absence of a definitive oracle for tax outcomes, coupled with the incompleteness of traditional tests, makes conventional metamorphic testing insufficient. 
To overcome this, we leverage higher-order metamorphic relations as a central solution. 
The key challenges, then, lie in \emph{automatically inferring these specifications} from tax regulations and generating comprehensive test cases that effectively exercise them.  

To realize this approach, we introduce a \emph{multi-agent system} that automates specification inference, code generation, and validation. 
Our system consists of five specialized agents, each with a distinct role in the pipeline: a \emph{Tax Expert Agent}, two \emph{Coder Agents}, a \emph{Senior Coder Agent}, and a \emph{Metamorphic Agent}. 
Figure~\ref{fig:framework} illustrates how these agents interact, highlighting their collaboration in synthesizing and verifying tax software against legal requirements.

\subsection{Individual Agents and Division of Labor}
The overarching goal of our multi-agent system is to autonomously generate, validate, and refine tax calculation functions to ensure accuracy and compliance with tax rules. 
Each agent contributes specialized capabilities toward this objective. 
By assigning distinct roles, we can leverage the strengths of different LLMs and promote a more efficient, robust development process.

\subsubsection{TaxExpertAgent}
This agent serves as the initial interpreter of the legal text. 
It parses tax regulations and converts them into a structured JSON representation. 
It also produces JSON-based specifications for individual functions required in tax software, describing their purpose, inputs, outputs, calculations, and edge cases. 
The output JSON must conform to a predefined schema that validates data types (e.g., tax rates as numbers), required fields, and structural integrity (e.g., verifying that \texttt{tax\_brackets} for a filing status is an array of objects, each with a \texttt{threshold} and a \texttt{rate}).  

If the generated JSON fails schema validation, the agent regenerates it until compliance is achieved. 
For example, the JSON structure for tax brackets may include keys for filing statuses, each containing bracket thresholds and corresponding rates. 
A JSON output from \texttt{TaxExpertAgent} for bracket calculation is shown below:
\begin{lstlisting}
{
  "tax_brackets": {
    "single": [
      { "threshold_amount": 9950, "rate_decimal": 0.1 },
      { "threshold_amount": 40525, "rate_decimal": 0.12 },
      { "threshold_amount": 86375, "rate_decimal": 0.22 }
    ],
    "married_filing_jointly": [...]
  },
  "standard_deductions": {
    "single": {
      "base_amount": 12400,
      "additional_elderly": 1650,
      "additional_blind": 1650
    },
    ...
  }
}
\end{lstlisting}

\subsubsection{Senior Coder and Coder Agents}
Within our framework, the \texttt{CoderAgent} and \texttt{SeniorCoderAgent} collaborate to generate the Python code for tax functions. 
Guided by JSON-based specifications from the \texttt{TaxExpertAgent}, the \texttt{CoderAgent} instances produce code that adheres to defined inputs, outputs, constraints, and edge cases.  
The two \texttt{CoderAgent} instances participate in an internal review and refinement cycle orchestrated by the \texttt{SeniorCoderAgent}. 
First, one \texttt{CoderAgent} generates an initial implementation, and then the \texttt{SeniorCoderAgent} evaluates this code against the specification.
If the implementation is satisfactory, it is accepted; otherwise, the \texttt{SeniorCoderAgent} provides feedback, and the second \texttt{CoderAgent} generates a revised version incorporating these corrections.  
Although both \texttt{CoderAgent} instances use the same base LLM, we apply slightly different temperature settings (e.g., a 0.1--0.2 variation). 
This encourages diverse outputs and exploration of alternatives without destabilizing the process.

\subsubsection*{Function Description (Example):}
Each \texttt{CoderAgent} receives detailed instructions through JSON-formatted function descriptions, ensuring adherence to specifications and avoiding hardcoded values by referencing tax rules via JSON keys.  An example of such a description for the \texttt{calculate\_tax} function is provided next:
\begin{lstlisting}
{
  "function_name": "calculate_tax",
  "inputs": {
    "income": { "type": "float", "constraints": ["income >= 0"] },
    "status": { "type": "string", "constraints": [
      "in ['single', 'married_filing_jointly']"
    ]}
  },
  "outputs": {
    "tax_due": { "type": "float", "rounding": 2 }
  },
  "calculations": [
    { "step": "apply_brackets", "formula": "income * t_rate", ... }
  ]
  ...
}
\end{lstlisting}
This structured JSON input ensures that \texttt{CoderAgent} implementations remain accurate, consistent, and aligned with tax rules. 

\begin{figure}[t!]
    \centering
    \includegraphics[width=0.95\linewidth, trim=0 0 0 51pt, clip]{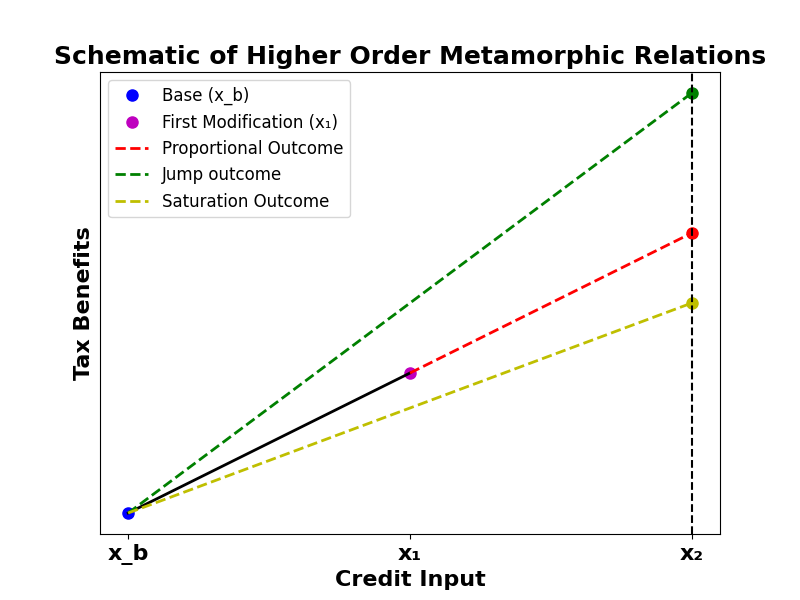}
     \caption{Schematic of Higher Order Metamorphic Relations }
     \Description{Chart with horizontal axis labeled Credit Input and vertical axis labeled Tax Benefits. It shows a solid black line establishing a baseline from the base input to the first modification. From the second modification, three dashed curves diverge: a red proportional increase, a green threshold jump, and a yellow saturation outcome.}
    \label{fig:hmtchart}
\end{figure}
\subsection{Higher-Order Metamorphic Testing Agent}
\label{sec:approach}
Integrated into the sequential workflow, the \texttt{MetamorphicAgent} receives generated code from the \texttt{SeniorCoderAgent} and subjects it to rigorous metamorphic testing. Previously, Tizpaz-Niari et al.~\cite{ICSE-SEIS23} manually extracted metamorphic relations from the legal tax documents. 
Srinivas et al.~\cite{srinivas2023potential} presented a few-shot in-context learning to automatically infer metamorphic specifications from legal documents.
Following this literature, we initially implement basic metamorphic testing, perturbing single inputs (e.g., income) and verifying that the resulting change in the output aligns with expected directional relationships (e.g., increased income should lead to increased tax). 
However, we face two key challenges: i) generating metamorphic specifications via LLMs in the language of first-order logic is challenging, even for 4-ary relations, and ii) the directional oracle misses many existing bugs.

The \texttt{MetamorphicAgent} is engineered to analyze the \textit{rate of change} in tax calculations across multiple, related input variations, not just pairwise comparisons of the directional change. 
This allows us to encode more precise relationships and increase the soundness of our code generation. The agent considers three categories for metamorphic relations: i) \textit{proportional increase} that verifies proportional changes in tax output for incremental input changes; ii) \textit{threshold jump} that tests expected ``jumps'' in tax rate change when inputs cross critical thresholds; and iii) \textit{saturation} that validates tax outputs remain invariant when inputs are within saturation ranges.

\begin{figure}[t!]
\centering
\begin{lrbox}{\grayframebox}
  \begin{minipage}{\dimexpr\linewidth-2\fboxsep-2\fboxrule\relax}
    \small
    \textcolor{blue!60!black}{\textbf{Goal:}} Suggest input tuples for \texttt{income} for a \textbf{Threshold Jump} test.

    \vspace{0.4em}
    \textcolor{blue!60!black}{\textbf{Understanding the Test:}}
    Verify that the software correctly handles discrete jumps in the rate of change when an input crosses a tax-rule boundary (e.g., a tax bracket).

    \vspace{0.4em}
    \textcolor{blue!60!black}{\textbf{Key Requirements for Tuples ($x_b, x_1, x_2$):}}
    \begin{itemize}
      \item \textbf{Threshold Targeting:} Identify a specific tax bracket boundary from the provided legal text.
      \item \textbf{Values Around Threshold:} Place $x_b$ and $x_1$ just \textit{below} the threshold, and $x_2$ just \textit{above}, ensuring a crossing between $x_1$ and $x_2$.
      \item \textbf{Meaningful Jumps:} Ensure the increment from $x_1$ to $x_2$ is large enough to trigger a distinct change in the tax output.
    \end{itemize}

    \vspace{0.4em}
    \textcolor{blue!60!black}{\textbf{Task:}} Suggest at least four tuples testing different bracket jumps. Respond with a valid JSON containing a list of tuples.

    \vspace{0.4em}
    \textcolor{blue!60!black}{\textbf{Example Response Format:}}

\begin{lstlisting}
{
  "suggested_tuples": [
    {
      "input_tuple": [35000, 40525, 48000],
      "reason": "Testing 12% to 22% bracket jump"
    },
    ...
  ]
}
\end{lstlisting}
  \end{minipage}
\end{lrbox}

\noindent\fcolorbox{gray!50}{gray!5}{\usebox{\grayframebox}}

\caption{Summary of the HMT prompt provided to the \texttt{MetamorphicAgent} for the \texttt{ThresholdJump} test category.}
\Description{A framed box showing a worked example prompt for a Threshold Jump test on income. 
It outlines the goal of suggesting input tuples, explains that the test verifies correct handling of discrete jumps at tax bracket boundaries, lists requirements for tuples (targeting a threshold, placing values just below and above it, and ensuring meaningful jumps),  and ends with a JSON example format that specifies income tuples and reasons.}
\label{fig:hmt-prompt-example}
\end{figure}

Figure~\ref{fig:hmtchart} illustrates our higher-order metamorphic relations. In this chart, the horizontal axis (``Credit Input'') marks the specific input that is being tested: the base input ($x_b$), the first modification ($x_1$), and the second modification ($x_2$). The black solid line from $x_b$ to $x_1$ establishes the baseline rate of change. At $x_2$, the chart shows three possible (expected) outcomes for the tax benefit: the red dashed line represents a \textit{proportional increase}---indicating that the tax benefit changes at the same rate as the baseline; the green dashed line indicates a \textit{threshold jump} where the tax benefit increases more steeply once a critical input threshold is exceeded; and the yellow dashed line depicts \textit{saturation}, where the tax benefit remains nearly invariant despite further increases in input.

Instead of manually encoding metamorphic relations in first-order logic, we instruct an LLM-based \texttt{MetamorphicAgent} to generate test cases. Using natural language descriptions, tax rules, illustrative examples, and the required output format, the agent generates diverse test cases for the given input category. When it detects a metamorphic relation violation—indicating unexpected tax behavior—it provides concrete counterexamples to iteratively enhance the reliability and correctness of the tax software.

\subsubsection*{Example: Publication 970}. The American Opportunity Tax Credit (AOTC), outlined in IRS Publication 970, follows a tiered structure to incentivize the first \$4,000 of educational expenses. Specifically, it covers 100\% of the first \$2,000 and 25\% of the next \$2,000, with a maximum credit of \$2,500. A basic 4-ary metamorphic relation may only check whether increasing qualified expenses leads to a higher credit. However, a higher-order relation can examine the \textit{rate of increase} in credit as expenses vary. 

The \texttt{MetamorphicAgent} can verify whether the \textit{marginal increase} in credit is significantly higher for expenses in the \$0-\$2,000 range (100\%) compared to the \$2,001-\$4,000 range (25\%). Additionally, it can ensure that the credit increase ceases once expenses exceed \$4,000, as the credit cap is reached. By analyzing these changing rates of credit increase for the label \(\{qualified\_expenses\}\), the agent enables more precise validation and error detection.

\begin{lstlisting}
"discrepancies": [
  { 
    "input": "qualified_expenses",
    "test_category": "ProportionalIncrease",
    "filing_status": "single",
    "base_value": 1000,
    "new_value_1": 2000,
    "new_value_2": 3000,
    "verification_result": "FAIL",
    "verification_reason": "Credit rate increase isn't consistent with AOTC's tiered structure. The rate should decrease after $2000 expenses, but it remains proportionally same.",
    "initial_tax": 1000.00,
    "modified_tax_tuple": [ 7400.00, 6200.00, 5000.00 ],
    "Rate_change_base (R1)": 1.20,
    "Rate_change_follow-up (R2)": 1.20
  }
]
\end{lstlisting}

\subsubsection*{Worked Example: Generating a Threshold Jump Test.}  
To illustrate how the \texttt{MetamorphicAgent} generates higher-order test cases, we present a worked example for the \textbf{Threshold Jump} category targeting the \texttt{income} input. 
The objective is to verify that tax calculation reflects the discrete jump in marginal rates when income crosses a bracket boundary, a defining feature of progressive taxation.

The agent receives a detailed prompt that first explains the metamorphic property being tested and then provides instructions for generating valid inputs. 
A summary of this prompt is shown in Figure~\ref{fig:hmt-prompt-example}. 
Specifically, the LLM is asked to identify a tax threshold from the legal context (e.g., the boundary between the 12\% and 22\% brackets at \$40{,}525 for a single filer) and generate a tuple of three income values ($x_b, x_1, x_2$) positioned to straddle this threshold.

Given this prompt, a capable LLM might return the tuple (\texttt{35000, 40525, 48000}) for a single filer. 
The test then compares two rates of change relative to the baseline $x_b$:
\begin{itemize}
    \item The first rate of change, $R_1$, is computed between the baseline $x_b$ and the first modification $x_1$, which lies at the bracket boundary. 
    Since both points are effectively within the 12\% marginal tax bracket, the rate is
    \[
    R_1 = \frac{F(x_1) - F(x_b)}{x_1 - x_b} \approx 0.12.
    \]
    \item The second rate of change, $R_2$, is computed between the same baseline $x_b$ and the second modification $x_2$, which crosses into the 22\% bracket. 
    This rate reflects the average change over a range that includes income taxed at 22\% rate:
    \[
    R_2 = \frac{F(x_2) - F(x_b)}{x_2 - x_b}.
    \]
\end{itemize}
The metamorphic test passes if the implementation shows $R_2 {>} R_1$, confirming that the average tax rate over the wider range ($x_b$ to $x_2$) is higher than over the initial range ($x_b$ to $x_1$), as expected in a progressive system. 
A faulty implementation, such as applying a flat rate, would fail this test because $R_1$ and $R_2$ would be approximately equal.
As a sanity check, we manually reviewed a subset of generated test tuples during development to ensure they were logically sound (e.g., values straddled meaningful thresholds).

\subsection{Multi-Agent Collaboration}
Our framework employs a team of specialized agents, each contributing distinct capabilities to the tax code generation process. 
Below, we outline their collaborative interactions.

\subsubsection*{Code Generation.}  
The workflow begins with the \texttt{TaxExpertAgent}, which produces tax rule data in JSON format for specific scenarios. 
This JSON is refined until it satisfies schema requirements. 
Once it is validated, the \texttt{TaxExpertAgent} generates JSON-based function specifications describing each required function’s purpose, inputs, outputs, edge cases, and computational steps. 
These serve as blueprints for the \texttt{SeniorCoderAgent}, which coordinates multiple \texttt{CoderAgent} instances to implement code that accesses values from the tax-rule JSON rather than hardcoding constants. 
This design simplifies updates, improves debugging, and enforces consistency. 
The \texttt{SeniorCoderAgent} then selects the best candidate implementations from the coders.

\subsubsection*{Metamorphic Testing.} 
The \texttt{MetamorphicAgent} serves as the driver of our counterexample-guided refinement process, working closely with the \texttt{SeniorCoderAgent}.
It applies higher-order metamorphic relations derived from tax law descriptions to automatically generate test cases and detect discrepancies---violations of expected tax behavior. 
For each discrepancy, it reports the triggering inputs, outputs, and implicated functions. 
The \texttt{SeniorCoderAgent} then uses these counterexamples to repair the corresponding functions. 
Through this iterative feedback loop, guided by higher-order testing, the framework steadily improves both the robustness and functional correctness of the generated code.

Our results reflect the outcome of a single run of the framework---from initial JSON specifications through code generation, metamorphic testing, and refinement by the \texttt{SeniorCoderAgent}.

\section{Experiments}
\label{sec:experiments}

We conduct a comprehensive evaluation of our multi-agent framework for tax software generation. 
Our experiments are designed to measure how well LLMs, both standalone and embedded in our agentic system, can translate U.S. federal tax law into executable code. 
We benchmark performance across six progressively complex tax scenarios derived from IRS publications, ranging from basic bracket computations to retirement distributions (1099-R). 

\subsection{Experimental Setup}
\subsubsection{Benchmarks.}
\label{sec:experiment-benchmark}
To evaluate the effectiveness of LLMs in the tax software domain, we design six benchmarks derived from the U.S. federal income tax code. 
These benchmarks progress in difficulty, from basic computations to complex multi-form scenarios:

\begin{enumerate}
    \item \textbf{Benchmark 1: Brackets and Standard Deductions.} 
    A foundational task requiring the calculation of income tax based on income and filing status, incorporating tax brackets, standard deductions, and age/blindness adjustments.  

    \item \textbf{Benchmark 2: Earned Income Tax Credit (EITC)~\cite{pub596}.}  
    Adds the refundable EITC, which supports low-to-moderate income workers and families. The LLM must compute credits using income, filing status, and number of children, while handling the EITC’s nonlinear phase-in and phase-out rules.  

    \item \textbf{Benchmark 3: Child Tax Credit (CTC) and Other Dependent Credit (ODC)~\cite{8812}.} Expanding the scope further, this benchmark includes the CTC and ODC, which provide tax benefits to families with children and other dependents. The LLMs must account for phase-out thresholds and credit amounts, which vary based on income and filing status.
    
    \item \textbf{Benchmark 4: American Opportunity Tax Credit~\cite{8863}.} This benchmark focuses on the AOTC, a tax credit for qualified education expenses. To accurately compute the credit amount, the LLMs must consider various factors, including income, filing status, qualified expenses, scholarships, enrollment status, and the student's year in school.
    
    \item \textbf{Benchmark 5: Itemized Deductions~\cite{1040sa}.} This benchmark introduces the complexity of itemized deductions, allowing taxpayers to deduct certain expenses from their Adjusted Gross Income (AGI) if they exceed the standard deduction. Key deductions include medical expenses that surpass 7.5\% of AGI, a capped state and local tax (SALT) deduction of \$10,000, as well as deductible home mortgage interest, charitable contributions, and casualty or theft losses. 

    \item \textbf{Benchmark 6: 1099-R Distributions and Penalties~\cite{1099-r}.} The most complex benchmark in our suite, Scenario 6, centers on retirement and pension distributions reported on IRS Form 1099-R. 
    It requires computing taxable amounts for distributions from diverse sources such as IRAs, annuities, and pensions. 
    Key challenges include handling capital gains, applying penalties for early withdrawals under specific conditions, and correctly processing exception codes that waive these penalties. 
    Accurate implementation also demands use of the IRS Simplified Method to determine the non-taxable portion of annuity distributions based on factors like cost basis and annuitant age. 
    Finally, the system must manage distribution codes in both string and numeric formats and apply conditional logic tailored to taxpayer circumstances.
\end{enumerate}

\subsubsection{Large Language Models.}
We evaluate the capabilities of several large language models (LLMs) in automating the process of generating and validating tax software from natural legal code:
\begin{enumerate}
\item \textbf{GPT-4o (OpenAI).} A large-scale language model with 200 billion parameters. Known for its advanced natural language understanding and robust code generation capabilities, it excels in complex tasks requiring nuanced language comprehension and precise syntactic structure, making it indispensable in generating precise tax-related functions.

\item \textbf{GPT-4o-mini (OpenAI).} A smaller and faster variant of GPT-4o, with 8 billion parameters. GPT-4o-mini offers improved computational efficiency, making it a suitable choice for tasks where speed is prioritized over the depth of language modeling. This distinction highlights GPT-4o-mini as a small-scale model in contrast to the large-scale GPT-4o.

\item \textbf{Claude 3.5-Sonnet (Anthropic).} A large language model developed by Anthropic, emphasizing safety and user alignment in language processing.

\item \textbf{Llama 3.1-70B (Meta).} An open-source model developed by Meta with 70 billion parameters, representing a large-scale LLM in the Llama family.

\item \textbf{Llama 3.1-8B (Meta).} A smaller variant in the Llama family with 8 billion parameters. Despite its smaller scale, Llama 3.1-8B is highly practical, as it can run on edge devices.
\end{enumerate}

\subsubsection{Prompting Approach.}
\label{sec:experiment-prompting}
To evaluate the capabilities of LLMs in translating tax code to tax software, we employ two promptings:
\begin{enumerate}
\item \textbf{Zero-Shot Prompting.} This generates code autonomously based on its general understanding of natural language, tax rules, and programming without any other context.

\item  \textbf{Step-Wise Chain-of-Thought (COT) Prompting.} In Step-Wise Chain-of-Thought approach~\cite{yu2023towards}, the LLM is guided through the task with structured interactions 
in two stages:
    \begin{enumerate}
        \item \textbf{Reasoning Stage.} The LLM is prompted to reason through the tax rules and specific tax code requirements. This stage encourages the model to identify key tax policies ---e.g., income thresholds, applicable deductions, and credit limits---and to organize these elements logically.
        
        \item \textbf{Code Generation Stage.} In the second interaction, the LLM uses its structured reasoning from the first stage to generate the necessary code. 
    \end{enumerate}
\end{enumerate}

\subsection{Evaluation Metric and Technical Details}
\label{sec:experiment-symbolic}
\subsubsection{Symbolic Executions.} To thoroughly evaluate the generated tax code, we apply symbolic execution techniques on the ground-truth tax software. 
By running symbolic execution on the ground-truth tax software, we collect test cases reflecting each tax benchmark's expected logic and outcomes. These generated test cases are then used to evaluate the generated code by comparing its outputs against the ground-truth software. 

\subsubsection{Ground-Truth Oracle \& Cross-Reference.}
We evaluate synthesized programs against a hand-authored reference implementation fixed to \textbf{Tax Year 2021} and aligned strictly with the rules described in our scenarios (\S4.1). The oracle spans \textbf{10 reference functions} covering: progressive bracket computation (by filing status), standard deductions including age/blind add-ons, EITC phase-in/plateau/phase-out, CTC/ODC thresholds, AOTC’s \(100\%\!/25\%\) tiers with a \$2{,}500 cap, itemized deductions (SALT cap; medical \(7.5\%\) AGI floor), and 1099-R taxable amounts, early-withdrawal penalties, and the IRS \emph{Simplified Method}. To ensure the correctness of the oracle implementations, we perform stress-testing at boundaries (bracket knees, phase-in/out breakpoints, and cap/saturation regions). Where scope overlapped with open-source tools (e.g., OpenTaxSolver~\cite{openTaxSolver}, Colorado Toolbox~\cite{ColoradoTaxAid}), we performed targeted spot-checks at boundaries to check the correctness of our oracle implementations. 

\subsubsection{Evaluation Metric.}
\label{sec:evaluation-metric}
In this study, we considered two common evaluation metrics for code generation: \textbf{Pass@k} and \textbf{Partial Pass@k}, which measure the success rate of generating correct solutions within $k$ attempts. \textbf{Pass@k} reflects the proportion of generations passing all test cases. However, because many benchmarks here involve complex requirements, most LLMs failed to pass all tests, making \textbf{Pass@k} an unsuitable measure of performance.

Instead, we chose \textbf{Partial Pass@k} as our primary metric, which measures the average percentage of successful test cases over $k$ generations of code. For a comprehensive assessment, we report the following variants:
\begin{itemize}
    \item \textbf{Partial Pass@10}: Average percentage of successful test cases across 10 generations, providing an overview of performance across attempts.
    \item \textbf{Partial Pass@1}: Highest percentage of test cases passed in a single generation, reflecting best-case performance.
    \item \textbf{Worst@10}: Lowest percentage of successful test cases across 10 generations, representing worst-case performance.
\end{itemize}

\subsubsection{Technical Details.}
\label{sec:technical-details}
Our experiments were conducted on an Amazon Web Services (AWS) \texttt{g5.8xlarge} Ubuntu server, equipped with GPU support. The framework used for implementing the agentic approach in this project was the \texttt{agentlite} library by Salesforce \cite{Liu2024AgentLiteAL}, which provided the necessary abstractions for orchestrating multi-agent interactions. Symbolic execution was carried out using the \texttt{Z3} solver~\cite{de2008z3} to generate and validate test cases based on the tax benchmarks. Temperature value for all the models is set to 0.5.

\subsection{Research Questions}
We investigate the following questions:

\begin{enumerate}[start=1,label={\bfseries RQ\arabic*},leftmargin=3em]

    \item How effectively do baseline LLMs, using zero-shot and step-wise chain-of-thought prompting, perform in generating tax software code across scenarios of varying complexity?

    \item How does our LLM-based multi-agent design compare to the baseline LLMs? 

    \item How does the integration of the metamorphic testing agent contribute to the correctness of the generated tax software?

    \item What is the contribution of each agent in the collaborations? and what are the cost-accuracy trade-offs?

\end{enumerate}

\subsubsection{RQ1. Baseline LLM Techniques.}
\label{sec:RQ1}
In examining RQ1, we evaluate the baseline performance of large language models (LLMs) in generating tax software directly from tax code across two prompting methods: \textit{Zero-Shot} and \textit{Step-Wise Chain-of-Thought (CoT)}. 

\begin{table}[t!]
\centering
\caption{Zero shot prompting results}

\resizebox{0.47\textwidth}{!}{ 
\begin{tabular}{|c|l|l|l|l|l|l|l|}
\hline
\multicolumn{1}{|l|}{Model} & Metric & Scen.1 & Scen.2 & Scen.3 & Scen.4 & Scen.5 & Scen.6 \\ \hline
\multirow{3}{*}{llama 8b} & PP@1 & 53\% & 0\% & 0\% & 0\% & 15\% & 0\% \\ \cline{2-8} 
 & PP@10 & 19\% & 0\% & 0\% & 0\% & 3\% & 0\% \\ \cline{2-8} 
 & worst@10 & 4\% & 0\% & 0\% & 0\% & 0\% & 0\% \\ \hline
\multirow{3}{*}{llama70b} & PP@1 & 100\% & 71\% & 92\% & 42\% & 18\% & 0\% \\ \cline{2-8} 
 & PP@10 & 51\% & 27\% & 46\% & 24\% & 5\% & 0\% \\ \cline{2-8} 
 & worst@10 & 32\% & 8\% & 15\% & 10\% & 0\% & 0\% \\ \hline
\multirow{3}{*}{gpt-4o-mini} & PP@1 & 100\% & 83\% & 12\% & 23\% & 32\% & 4\% \\ \cline{2-8} 
 & PP@10 & 88\% & 40\% & 8\% & 6\% & 10\% & 2\% \\ \cline{2-8} 
 & worst@10 & 62\% & 10\% & 0\% & 0\% & 0\% & 0\% \\ \hline
\multirow{3}{*}{gpt-4o} & PP@1 & 100\% & 91\% & 94\% & 98\% & 98\% & 23\% \\ \cline{2-8} 
 & PP@10 & 95\% & 82\% & 71\% & 62\% & 56\% & 18\% \\ \cline{2-8} 
 & worst@10 & 88\% & 65\% & 37\% & 18\% & 27\% & 5\% \\ \hline
\multirow{3}{*}{claude 3.5} & PP@1 & 100\% & 100\% & 96\% & 97\% & 98\% & 39\% \\ \cline{2-8} 
 & PP@10 & 100\% & 86\% & 96\% & 97\% & 93\% & 29\% \\ \cline{2-8} 
 & worst@10 & 100\% & 72\% & 96\% & 97\% & 84\% & 14\% \\ \hline
\end{tabular}
}
\label{tab:rq1}
\end{table}

\subsubsection*{Zero-Shot Results (Table \ref{tab:rq1}).}
Under the zero-shot prompting approach, where models receive minimal guidance, we observe clear trends in performance based on model size and task complexity:
\begin{itemize}
    \item \textbf{Superior Performance of Large Models in Simple Scenarios.} Large models such as \texttt{GPT-4o} (200B parameters) and \texttt{Claude 3.5} (175B parameters) achieve high partial pass rates in simpler benchmarks. Both models reach PP@1 scores of over 95\% in Scenarios 1 through 5.
    \item \textbf{Sharp Declines in Complex Scenarios.} As the benchmark complexity increases, most models show a drop in performance, with Scenario 6 (1099-R distributions) being particularly challenging. For example, \texttt{GPT-4o} achieves a PP@1 score of 23\% and PP@10 of 18\%, while \texttt{Claude 3.5} achieves PP@1 of 39\% and PP@10 of 29\%. These results suggest that zero-shot prompting alone may be insufficient for handling multi-step tax rules involving multiple dependencies and conditional calculations.
    \item \textbf{Limited Capability of Smaller Models in All Scenarios.} Small models such as \texttt{Llama 8B} and \texttt{GPT-4o-mini} (8B parameters) struggled across benchmarks, showing low partial pass rates in all cases except the simplest one. 
\end{itemize}

\begin{table}[t!]
\centering
\caption{Step-Wise Chain of though prompting results}
\resizebox{0.47\textwidth}{!}{ 
\begin{tabular}{|c|l|l|l|l|l|l|l|}
\hline
\multicolumn{1}{|l|}{Model} & Metric & Scen.1 & Scen.2 & Scen.3 & Scen.4 & Scen.5 & Scen.6 \\ \hline
\multirow{3}{*}{llama 8b} & PP@1 & 59\% & 0\% & 0\% & 0\% & 0\% & 0\% \\ \cline{2-8} 
 & PP@10 & 31\% & 0\% & 0\% & 0\% & 0\% & 0\% \\ \cline{2-8} 
 & worst@10 & 12\% & 0\% & 0\% & 0\% & 0\% & 0\% \\ \hline
\multirow{3}{*}{llama70b} & PP@1 & 100\% & 98\% & 98\% & 28\% & 32\% & 0\% \\ \cline{2-8} 
 & PP@10 & 78\% & 47\% & 71\% & 12\% & 11\% & 0\% \\ \cline{2-8} 
 & worst@10 & 56\% & 10\% & 11\% & 0\% & 0\% & 0\% \\ \hline
\multirow{3}{*}{gpt-4o-mini} & PP@1 & 100\% & 82\% & 10\% & 5\% & 2\% & 21\% \\ \cline{2-8} 
 & PP@10 & 84\% & 46\% & 6\% & 2\% & 1\% & 9\% \\ \cline{2-8} 
 & worst@10 & 72\% & 0\% & 3\% & 0\% & 0\% & 0\% \\ \hline
\multirow{3}{*}{gpt-4o} & PP@1 & 100\% & 98\% & 93\% & 68\% & 64\% & 28\% \\ \cline{2-8} 
 & PP@10 & 96\% & 84\% & 78\% & 53\% & 53\% & 14\% \\ \cline{2-8} 
 & worst@10 & 91\% & 77\% & 67\% & 41\% & 27\% & 9\% \\ \hline
\multirow{3}{*}{claude 3.5} & PP@1 & 100\% & 100\% & 98\% & 99\% & 98\% & 42\% \\ \cline{2-8} 
 & PP@10 & 100\% & 94\% & 98\% & 98\% & 98\% & 31\% \\ \cline{2-8} 
 & worst@10 & 100\% & 85\% & 98\% & 96\% & 98\% & 15\% \\ \hline
\end{tabular}
}
\label{tab:rq12}
\end{table}

\subsubsection*{Step-Wise Chain-of-Thought Results (Table \ref{tab:rq12}).}
The chain-of-thought (CoT) prompting yields substantial improvements in both accuracy and consistency, particularly for large models:
\begin{itemize}
    \item \textbf{Enhanced Consistency and Accuracy in Large Models.} CoT prompting boosts accuracy by a small margin, but it improves consistency of models, especially in moderately complex benchmarks. For instance, \texttt{GPT-4o} achieves worst@10 scores of 77\% in Scenario 2, up from 65\% in zero-shot.
    
    \item \textbf{Top-Performing Model in Complex Scenarios.} \texttt{Claude 3.5} shows the highest scores across complex benchmarks, with near-perfect PP@10 scores in Scenario 5 (98\%) and substantial improvements in Scenario 6, reaching a PP@1 score of 42\% and a PP@10 of 31\%.
    
    \item \textbf{Small Models Show Limited Benefit from CoT in Complex Scenarios.} CoT prompting provides only marginal improvements for smaller models, such as \texttt{Llama 8B} and \texttt{GPT-4o-mini}. For example, \texttt{GPT-4o-mini} shows minimal gains in Scenario 6 with a PP@10 of just 9\%, indicating that CoT prompting alone does not significantly enhance smaller models' abilities to handle complex tax logic.
\end{itemize}

\begin{answerbox}
\textbf{Answer RQ1}: Claude 3.5 with CoT prompting outperforms other models. 
While it achieves an accuracy up to 98\% in benchmarks with low-to-moderate complexity, 
it reaches PP@1 and PP@10 scores of 42\% and 31\% in the most challenging benchmark.
\end{answerbox}

\subsubsection{RQ2. Agentic Design for Tax Software.}
\label{sec:RQ2}
We evaluate our agentic framework \tool{} with three underlying language models: 1) \texttt{GPT-4o-mini} (8B), 2) \texttt{GPT-4o}, and 3) \texttt{Claude-3.5-Sonnet}.
As \texttt{claude-3.5 -sonnet} demonstrated the strongest baseline performance in our initial zero-shot and chain-of-thought experiments, we aimed to assess if the agentic framework could further amplify its capabilities. We also consider \texttt{GPT-4o-mini} and its larger variants \texttt{GPT-4o} to test the agentic framework with a range of models and demonstrate its effectiveness across diverse LLM capabilities. 
The results, detailed in Table~\ref{tab:rq2}, illustrate the effectiveness of our agentic approach, which we designed to enable a smaller model to achieve accuracy close to and, in some cases, surpassing that of larger, frontier models when using baseline prompting techniques.

\begin{table}[!tbh]
\centering
\caption{Performance of the Agentic framework}
\resizebox{0.47\textwidth}{!}{
\begin{tabular}{|cc|l|l|l|l|l|l|l|}
\hline
\multicolumn{1}{|l|}{Agents} & \multicolumn{1}{l|}{Model} & Metric & Scen.1 & Scen.2 & Scen.3 & Scen.4 & Scen.5 & Scen.6 \\ \hline
\multicolumn{1}{|c|}{\multirow{9}{*}{\begin{tabular}[c]{@{}c@{}}Coder\\  +S-Coder\\  +T-Expert\end{tabular}}} & \multirow{3}{*}{gpt-4o-mini} & PP@1 & 100\% & 100\% & 97\% & 94\% & 89\% & 78\% \\ \cline{3-9}
\multicolumn{1}{|c|}{} &  & PP@10 & 100\% & 100\% & 92\% & 84\% & 75\% & 62\% \\ \cline{3-9}
\multicolumn{1}{|c|}{} &  & worst@10 & 100\% & 100\% & 88\% & 75\% & 61\% & 45\% \\ \cline{2-9}
\multicolumn{1}{|c|}{} & \multirow{3}{*}{gpt-4o} & PP@1 & 100\% & 100\% & 100\% & 100\% & 95\% & 89\% \\ \cline{3-9}
\multicolumn{1}{|c|}{} &  & PP@10 & 100\% & 100\% & 97\% & 94\% & 91\% & 83\% \\ \cline{3-9}
\multicolumn{1}{|c|}{} &  & worst@10 & 100\% & 100\% & 93\% & 89\% & 84\% & 72\% \\ \cline{2-9}
\multicolumn{1}{|c|}{} & \multirow{3}{*}{claude-3.5} & PP@1 & 100\% & 100\% & 100\% & 100\% & 100\% & 93\% \\ \cline{3-9}
\multicolumn{1}{|c|}{} &  & PP@10 & 100\% & 100\% & 100\% & 100\% & 98\% & 85\% \\ \cline{3-9}
\multicolumn{1}{|c|}{} &  & worst@10 & 100\% & 100\% & 100\% & 100\% & 95\% & 78\% \\ \hline
\end{tabular}
}
\label{tab:rq2}
\end{table}

\subsubsection*{High Performance Across Scenarios.}
The agentic approach consistently delivers high performance across all benchmarks. For simpler scenarios (1 \& 2), \texttt{GPT-4o-mini}, \texttt{GPT-4o}, and \texttt{claude-3.5} all reach 100\% on both PP@1 and PP@10.  
In more complex benchmarks, such as Scenario 6, \texttt{claude-3.5} achieved PP@1, PP@10, and worst@10 scores of 93\%, 85\%, and 78\%, respectively.

\subsubsection*{Improvement of the Worst-case Accuracy.}
A key advantage of the agentic approach is its capacity to maintain consistency in accuracy. For example, \texttt{GPT-4o-mini} achieves a worst@10 score of 45\% in Scenario 6, a significant improvement compared to 9\% for CoT results. Similarly, \texttt{claude-3.5-sonnet} achieves a significantly higher worst@10 score of 78\% in Scenario 6.

\subsubsection*{Enabling Small Models to Handle Complex Reasoning Tasks.}
In zero-shot and CoT settings, smaller models (e.g., \texttt{GPT-4o-mini} and \texttt{Llama 8B}) often failed to pass any test cases in complex benchmarks, with PP@1 and worst@10 scores close to 0\%. By contrast, the agentic \texttt{GPT-4o-mini} achieved PP@10 scores of 75\% and 62\% in Scenario 5 and 6, respectively. 

\begin{answerbox}
\textbf{Answer RQ2:} Our multi-agent design with a smaller model like GPT-4o-mini shows performance similar to or better than frontier baseline models. With \texttt{claude-3.5} agents, our framework achieved PP@1, PP@10, and worst@10 scores of 93\%, 85\%, and 78\%, respectively, in the most complex scenario. 
\end{answerbox}

\begin{table*}[!tbh]
\centering
\caption{Our framework with MetamorphicAgent. MT and HMT refer to regular and higher-order metamorphic testing.}
\Description{Table comparing model performance across agents, metrics, and scenarios with and without metamorphic testing.}
\label{tab:rq3}
\resizebox{0.75\textwidth}{!}{%
\begin{tabular}{|c|l|l|l|l|l|l|l|l|}
\hline
\multicolumn{1}{|c|}{{Agents}} & \multicolumn{1}{c|}{{Model}} & \multicolumn{1}{c|}{{Metric}} & \multicolumn{1}{c|}{{Scen.1}} & \multicolumn{1}{c|}{{Scen.2}} & \multicolumn{1}{c|}{{Scen.3}} & \multicolumn{1}{c|}{{Scen.4}} & \multicolumn{1}{c|}{{Scen.5}} & \multicolumn{1}{c|}{{Scen.6}} \\ \hline
\multirow{9}{*}{Coder + S-Coder + T-Expert} & \multirow{3}{*}{gpt-4o-mini} & PP@1 & 100\% & 100\% & 97\% & 94\% & 89\% & 78\% \\ \cline{3-9}
 &  & PP@10 & 100\% & 100\% & 92\% & 84\% & 75\% & 62\% \\ \cline{3-9}
 &  & worst@10 & 100\% & 100\% & 88\% & 75\% & 61\% & 45\% \\ \cline{2-9}
 & \multirow{3}{*}{gpt-4o} & PP@1 & 100\% & 100\% & 100\% & 100\% & 95\% & 89\% \\ \cline{3-9}
 &  & PP@10 & 100\% & 100\% & 97\% & 94\% & 91\% & 83\% \\ \cline{3-9}
 &  & worst@10 & 100\% & 100\% & 93\% & 89\% & 84\% & 72\% \\ \cline{2-9}
 & \multirow{3}{*}{claude-3.5-sonnet} & PP@1 & 100\% & 100\% & 100\% & 100\% & 100\% & 93\% \\ \cline{3-9}
 &  & PP@10 & 100\% & 100\% & 100\% & 100\% & 98\% & 85\% \\ \cline{3-9}
 &  & worst@10 & 100\% & 100\% & 100\% & 100\% & 95\% & 78\% \\ \hline
\multirow{9}{*}{Coder + S-Coder + T-Expert + MT} & \multirow{3}{*}{gpt-4o-mini} & PP@1 & 100\% & 100\% & 96\% & 94\% & 88\% & 80\% \\ \cline{3-9}
 &  & PP@10 & 100\% & 100\% & 94\% & 88\% & 83\% & 68\% \\ \cline{3-9}
 &  & worst@10 & 100\% & 100\% & 91\% & 83\% & 78\% & 55\% \\ \cline{2-9}
 & \multirow{3}{*}{gpt-4o} & PP@1 & 100\% & 100\% & 100\% & 100\% & 98\% & 91\% \\ \cline{3-9}
 &  & PP@10 & 100\% & 100\% & 97\% & 94\% & 92\% & 86\%  \\ \cline{3-9}
 &  & worst@10 & 100\% & 100\% & 94\% & 92\% & 89\% & 83\% \\ \cline{2-9}
 & \multirow{3}{*}{claude-3.5-sonnet} & PP@1 & 100\% & 100\% & 100\% & 100\% & 100\% & 93\% \\ \cline{3-9}
 &  & PP@10 & 100\% & 100\% & 100\% & 100\% & 99\% & 90\% \\ \cline{3-9}
 &  & worst@10 & 100\% & 100\% & 100\% & 100\% & 98\% & 86\% \\ \hline
\multirow{9}{*}{Coder + S-Coder + T-Expert + HMT} & \multirow{3}{*}{gpt-4o-mini} & PP@1 & 100\% & 100\% & 100\% & 94\% & 89\% & 81\% \\ \cline{3-9}
 &  & PP@10 & 100\% & 100\% & 95\% & 91\% & 86\% & 75\% \\ \cline{3-9}
 &  & worst@10 & 100\% & 100\% & 92\% & 89\% & 82\% & 69\% \\ \cline{2-9}
 & \multirow{3}{*}{gpt-4o} & PP@1 & 100\% & 100\% & 100\% & 100\% & 100\% & 100\% \\ \cline{3-9}
 &  & PP@10 & 100\% & 100\% & 100\% & 96\% & 95\% & 93\% \\ \cline{3-9}
 &  & worst@10 & 100\% & 100\% & 100\% & 94\% & 91\% & 88\% \\ \cline{2-9}
 & \multirow{3}{*}{claude-3.5-sonnet} & PP@1 & 100\% & 100\% & 100\% & 100\% & 100\% & 100\% \\ \cline{3-9}
 &  & PP@10 & 100\% & 100\% & 100\% & 100\% & 99\% & 95\% \\ \cline{3-9}
 &  & worst@10 & 100\% & 100\% & 100\% & 100\% & 98\% & 93\% \\ \hline
\end{tabular}
}
\end{table*}

\subsubsection{RQ3. Contributions of Metamorphic Agent}
\label{sec:RQ3}
Next, we integrate the \texttt{MetamorphicAgent} into our agentic framework. We analyze the contributions of two levels of metamorphic testing: 4-ary metamorphic testing (MT) and higher-order metamorphic testing (HMT).
Table~\ref{tab:rq3} presents the performance improvements achieved by incorporating these metamorphic testing approaches.

\subsubsection*{Metamorphic Testing} (MT). We first generate counterexamples by directional changes in tax calculations (e.g., increasing income should increase tax).
This shows notable improvements. For instance, in Scenario 6 with \texttt{GPT-4o-mini}, incorporating 4-ary MT improved the worst@10 score from 45\% (without MT) to 55\%.
Similarly, with \texttt{GPT-4o}, the worst@10 metric in Scenario 6 rose from 72\% to 83\% with 4-ary MT, demonstrating robustness even with a stronger LLM. PP@10 for all scenarios also increased 5-10\%.

\subsubsection*{Higher-order Metamorphic testing} (HMT). Building upon 4-ary MT, we introduced Higher-order Metamorphic Testing (HMT) incorporating n-ary metamorphic relations.  The performance gains with HMT are even more pronounced, especially in robustness metrics.  In Scenario 6 with \texttt{GPT-4o-mini}, HMT further improved the worst@10 score to 69\%. The most striking improvement is observed with \texttt{GPT-4o} in Scenario 6, where HMT pushes the worst@10 score to 88\% (MT achieves a score of 72\%). HMT enables the framework to generate a fully correct code even in the most complex scenario.

\subsubsection*{PP@10 and Convergence with Metamorphic Testing:} Both MT and HMT contribute to improve the PP@10 score. HMT consistently yields the highest PP@10 scores, especially in complex scenarios.  
For example, in Scenario 6 with \texttt{GPT-4o-mini}, PP@10 reaches 68\% with MT and further increases to 75\% with HMT.

\begin{answerbox}
\textbf{Answer RQ3:} 
While generating counterexamples via metamorphic testing (MT) improves the functional correctness of tax software, Higher-order MT provides superior gains, especially in complex scenarios. For example, HMT improves worst@10 scores by 14\% over the MT for \texttt{GPT-4o-mini} in Scenario 6.
\end{answerbox}

\begin{table*}[!tbh]
\centering
\caption{Comparative Results of Abolishing Agents with Varying LLM Capabilities.}
\resizebox{0.75\textwidth}{!}{
\begin{tabular}{|c|c|c|c|c|c|c|c|c|}
\hline
Agents & Model & Metric & Scenario 1 & Scenario 2 & Scenario 3 & Scenario 4 & Scenario 5 & Scenario 6 \\
\hline
\multirow{9}{*}{Coder Only}
& \multirow{3}{*}{gpt-4o-mini} & PP@1 & 100\% & 83\% & 12\% & 23\% & 32\% & 4\% \\
& & PP@10 & 88\% & 40\% & 8\% & 6\% & 10\% & 2\% \\
& & worst@10 & 62\% & 10\% & 0\% & 0\% & 0\% & 0\% \\ \cline{2-9}
& \multirow{3}{*}{gpt-4o} & PP@1 & 100\% & 91\% & 94\% & 98\% & 98\% & 23\% \\
& & PP@10 & 95\% & 82\% & 71\% & 62\% & 56\% & 18\% \\
& & worst@10 & 88\% & 65\% & 37\% & 18\% & 27\% & 5\% \\ \cline{2-9}
& \multirow{3}{*}{claude-3.5-sonnet} & PP@1 & 100\% & 100\% & 96\% & 97\% & 98\% & 39\% \\
& & PP@10 & 100\% & 86\% & 96\% & 97\% & 93\% & 29\% \\
& & worst@10 & 100\% & 72\% & 96\% & 97\% & 84\% & 14\% \\ \hline
\multirow{9}{*}{\shortstack{Coder\\+S-Coder\\+T-Expert}}
& \multirow{3}{*}{gpt-4o-mini} & PP@1 & 100\% & 100\% & 97\% & 94\% & 89\% & 78\% \\
& & PP@10 & 100\% & 100\% & 92\% & 84\% & 75\% & 62\% \\
& & worst@10 & 100\% & 100\% & 88\% & 75\% & 61\% & 45\% \\ \cline{2-9}
& \multirow{3}{*}{gpt-4o} & PP@1 & 100\% & 100\% & 100\% & 100\% & 95\% & 89\% \\
& & PP@10 & 100\% & 100\% & 97\% & 94\% & 91\% & 83\% \\
& & worst@10 & 100\% & 100\% & 93\% & 89\% & 84\% & 72\% \\ \cline{2-9}
& \multirow{3}{*}{claude-3.5-sonnet} & PP@1 & 100\% & 100\% & 100\% & 100\% & 100\% & 93\% \\
& & PP@10 & 100\% & 100\% & 100\% & 100\% & 98\% & 85\% \\
& & worst@10 & 100\% & 100\% & 100\% & 100\% & 95\% & 78\% \\ \hline

\multirow{9}{*}{\shortstack{Coder\\+S-Coder\\+T-Expert + MT}}
& \multirow{3}{*}{gpt-4o-mini} & PP@1 & 100\% & 100\% & 96\% & 94\% & 88\% & 80\% \\
& & PP@10 & 100\% & 100\% & 94\% & 88\% & 83\% & 68\% \\
& & worst@10 & 100\% & 100\% & 91\% & 83\% & 78\% & 55\% \\ \cline{2-9}
& \multirow{3}{*}{gpt-4o} & PP@1 & 100\% & 100\% & 100\% & 100\% & 98\% & 91\% \\
& & PP@10 & 100\% & 100\% & 97\% & 94\% & 92\% & 86\% \\
& & worst@10 & 100\% & 100\% & 94\% & 92\% & 89\% & 83\% \\ \cline{2-9}
& \multirow{3}{*}{claude-3.5-sonnet} & PP@1 & 100\% & 100\% & 100\% & 100\% & 100\% & 93\% \\
& & PP@10 & 100\% & 100\% & 100\% & 100\% & 99\% & 90\% \\
& & worst@10 & 100\% & 100\% & 100\% & 100\% & 98\% & 86\% \\ \hline
\multirow{9}{*}{\shortstack{Coder\\+S-Coder\\+T-Expert + HMT}}
& \multirow{3}{*}{gpt-4o-mini} & PP@1 & 100\% & 100\% & 100\% & 94\% & 89\% & 81\% \\
& & PP@10 & 100\% & 100\% & 95\% & 91\% & 86\% & 75\% \\
& & worst@10 & 100\% & 100\% & 92\% & 89\% & 82\% & 69\% \\ \cline{2-9}
& \multirow{3}{*}{gpt-4o} & PP@1 & 100\% & 100\% & 100\% & 100\% & 100\% & 100\% \\
& & PP@10 & 100\% & 100\% & 100\% & 96\% & 95\% & 93\% \\
& & worst@10 & 100\% & 100\% & 100\% & 94\% & 91\% & 88\% \\ \cline{2-9}
& \multirow{3}{*}{claude-3.5-sonnet} & PP@1 & 100\% & 100\% & 100\% & 100\% & 100\% & 100\% \\
& & PP@10 & 100\% & 100\% & 100\% & 100\% & 99\% & 95\% \\
& & worst@10 & 100\% & 100\% & 100\% & 100\% & 98\% & 93\% \\ \hline
\end{tabular}
}
\label{tab:rq4}
\end{table*}

\subsubsection{RQ4. Analysis of Different Agents.}
\label{sec:RQ4}
This section analyzes each agent's distinct and synergistic contributions within our multi-agent framework. Table~\ref{tab:rq4} summarizes the performance across various agents and LLM capabilities.

\subsubsection*{Agents' Contributions.} The \textit{Tax Expert Agent} consistently emerges as one of the most impactful individual agents.  As demonstrated in Table~\ref{tab:rq4}, adding the Tax Expert Agent to the "Coder Only" setup (with \texttt{GPT-4o-mini}) dramatically increased PP@1 scores from 12\% to 97\% for a medium complexity scenario. This agent's ability to structure tax rules and create precise function descriptions is needed for accurate code generation, particularly as scenario complexity increases.  Furthermore, leveraging a more powerful LLM like \texttt{GPT-4o} for the Tax Expert Agent while keeping other agents on \texttt{GPT-4o-mini} improves robustness (see Worst@10 scores).
    
 The integration of \textit{Metamorphic Testing Agents} provides a significant synergistic boost to the agentic framework, primarily enhancing robustness and reliability. Adding 4-ary MT demonstrably improves worst@10 scores, indicating increased resilience. For example, in Scenario 6, HMT increased \texttt{GPT-4o-mini}'s worst@10 69\% and \texttt{GPT-4o}'s worst@10  to an impressive 88\%. 

\subsubsection*{Token Usage and Computational Cost Breakdown:}  
\begin{figure}
    \centering
    \includegraphics[width=0.45\textwidth]{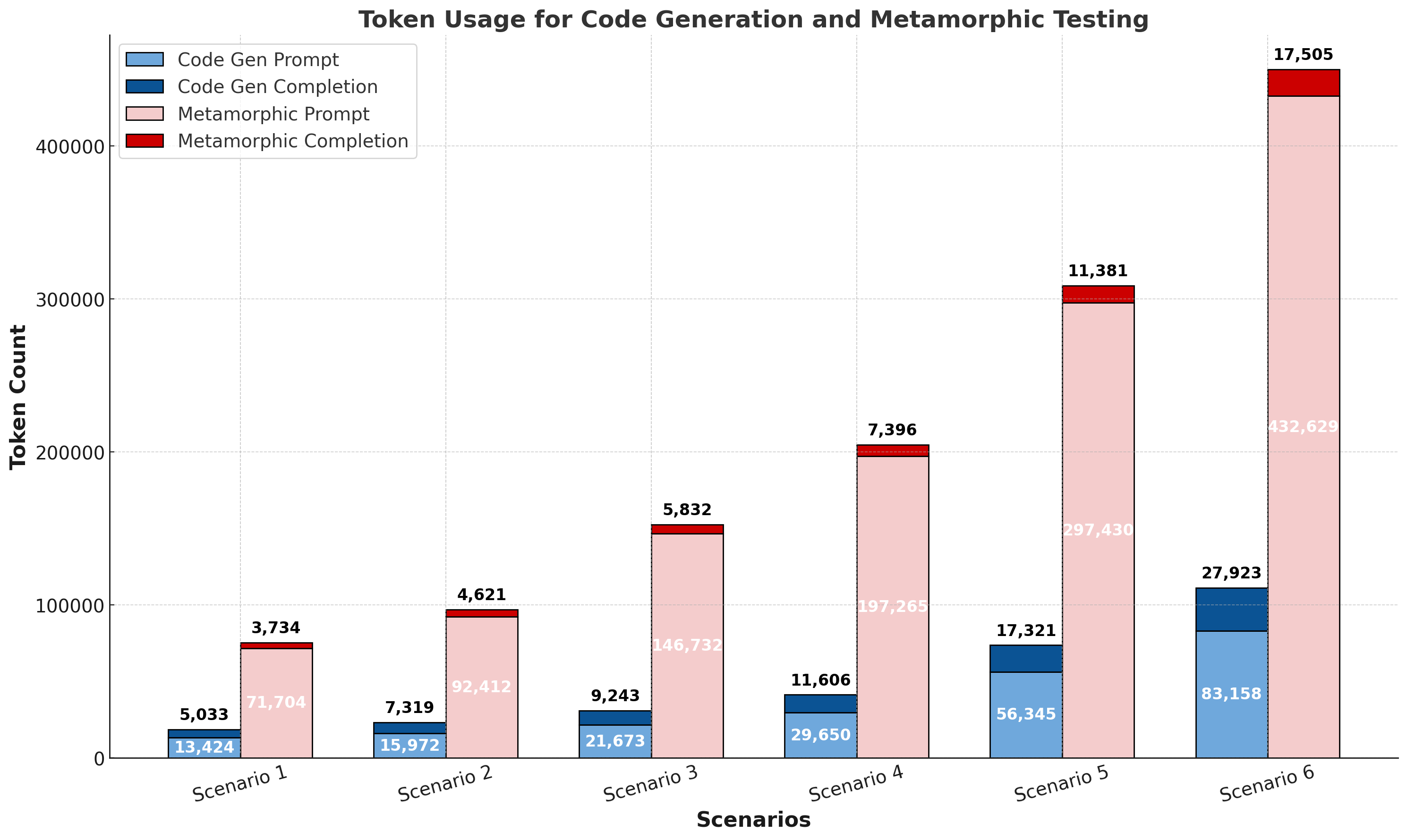}
    \caption{Token Usage: without MT, MT, and HMT.}
    \Description{Bar chart comparing token usage across scenarios under three conditions: no metamorphic testing, metamorphic testing, and higher-order metamorphic testing. Token consumption rises with scenario complexity, with large increases for MT and especially HMT.}
    \label{fig:tokens}
\end{figure}

A detailed token usage analysis reveals key computational cost considerations, particularly the balance between input and output tokens across different phases of our framework.  As depicted in Figure~\ref{fig:tokens}, token consumption increases with scenario complexity. For Scenario 1, basic agentic code generation totals 18,457 tokens, with metamorphic testing dramatically increasing this to 77,438 tokens, primarily due to a surge in input prompt tokens. This trend is amplified in Scenario 6, the most complex case, where code generation totals 111,081 tokens while incorporating Higher-Order Metamorphic Testing leads to a substantial increase to 450,134 tokens.  

\begin{answerbox}
\textbf{Answer RQ4:} Our results show that \texttt{TaxExpertAgent} and \texttt{MetamorphicTestingAgents} are critical components of our framework. 
Metamorphic Testing, especially HMT, significantly enhances robustness in the worst-case code generation. 
\end{answerbox}

\subsection{Discussion}
\label{sec:discussion}

\subsubsection*{Code-Specific LLMs.} We selected general-purpose LLMs over models specifically tuned for code generation. The rationale for selecting those models is driven by the need for models demonstrating expertise in natural legal language processing, code generation, and metamorphic test cases. 
While there are models specialized for code generation, the task at hand requires a deep understanding of tax regulations, often expressed in complex legal language.

\subsubsection*{Few-Shot Learning.} 
Few-shot prompting typically involves providing sample code snippets as examples for the LLM, guiding it to generate similar solutions. However, in the tax software domain, 
providing sample code may lead to outcomes that are more about pattern replication than an in-depth understanding of the tax logic. 

\subsubsection*{Agents with varying LLM capabilities.} 
While exploring all LLM combinations is infeasible, we perform a set of experiments to gain some insights into when agents are heterogeneous. For instance, in configurations where the \texttt{TaxExpertAgent} utilized \texttt{GPT-4o} while \texttt{CoderAgent} and \texttt{SeniorCoderAgent} used \texttt{GPT-4o-mini}, we observed a PP@10 score of 73\% in Scenario 6, compared to a score of 62\% for all \texttt{GPT-4o-mini} agents and 83\% for all  \texttt{GPT-4o} agents. This shows that a heterogeneous design may well bounded by smaller and larger homogeneous agents. We left further analysis in this direction to the future work. 

\subsubsection*{Legal-Critical Software.} Although we focused on the US-based tax prep software, the key takeaway is the application of the agentic approach with metamorphic testing for legal-critical software~\cite{10.1145/3306618.3314279,10.1145/3375627.3375807,10.1145/3392874}. For example,  our framework can be used to find errors in poverty management systems (e.g., the Pennsylvania ``Do I Qualify?''~\cite{COMPASS-HHS}) where the prior work relied on manual interpretation and assumed the presence of multiple versions of the same software~\cite{10.1145/3392874}. Our approach can automate policy translation into executable code and validate it with metamorphic testing. 

\subsubsection*{Limited Categories of Higher-Order MT} 
Despite HMT's improved bug detection capabilities over basic MT, our approach does not guarantee identifying and eliminating all potential errors. 
Furthermore, our HMT inherently focuses on three categories of metamorphic relations and does not guarantee that all possible tax calculation behaviors or edge cases will be covered.  

\subsubsection*{Ground-Truth Software.} Our evaluation significantly depends on the correctness of ground-truth tax software.
We manually check the correctness of our reference code by inspecting the source code, stress-testing at boundaries, and cross-referencing to open-source tax software toolkits. Due to the lack of formal guarantees, our reference implementations may contain logical bugs.

\section{Related Work}
\label{sec:related-work}
Research on multi-agent systems for automated code generation, debugging, and testing has made considerable progress. 

\subsubsection*{Agent-Based Code Generation.} Several studies focus on agent-based code generation and testing. For instance, AgentCoder~\cite{huang2023agentcoder} organizes agents to generate and refine code through iterative feedback, creating a modular structure for complex tasks. Similarly, CodeChain~\cite{le2024codechain} divides projects into smaller, manageable sub-modules refined through self-revisions, allowing for better integration to address intricate coding tasks. In a related approach, SoapFL~\cite{qin2024agentfl} includes agents dedicated to validation and fault identification, ensuring that issues are promptly identified. Recent frameworks such as ReAct~\cite{yao2023react} and Reflexion~\cite{shinn2023reflexion} emphasize the benefits of interleaving reasoning with action and self-reflection.

\subsubsection*{Feedback Mechanisms.} Feedback loops are crucial for refining code quality. SelfRefine~\cite{huang2023selfrefine} builds in feedback mechanisms that enable agents to iteratively analyze and improve their outputs (e.g., source-code generation), while CodePlan~\cite{bairi2023codeplan} incorporates planning capabilities to adjust actions based on previous changes adaptively.

\subsubsection*{Debugging and Fault Localization.} Debugging and fault localization are also critical. SoapFL~\cite{qin2024agentfl} utilizes specialized agents for project-level code review to identify errors through detailed inspection, and RCAgent~\cite{rcagent2024} adds memory components to track decisions, facilitating error localization in cloud systems.

\subsubsection*{Metamorphic Testing.} For testing without known correct answers, metamorphic testing is especially effective. Recent work such as METAL~\cite{hyun2024metal} can automate generation of metamorphic relations.

\subsubsection*{Collaborative Frameworks.} Collaborative frameworks boost the effectiveness of these systems. 
Gentopia~\cite{gentopia2023} organizes agents in a network where each insight contributes to  reliable outcomes. Other architectures~\cite{mapcoder2024, codes2024}, such as layered or tree-like structures, efficiently manage agent interactions. Unified approaches like RCAgent~\cite{rcagent2024} integrate error localization and solution generation into a collaborative process. 
\emph{The key innovation of our work lies in integrating legal experts into the multi-agent system design and introducing a metamorphic testing agent within the code generation process.}

\section{Conclusion}
\label{sec:conclusion}
This work introduced an agentic approach for generating reliable tax software from complex legal tax codes, with a central focus on metamorphic testing. Our multi-agent system, comprising specialized agents for tax expertise, code generation, and quality control, effectively interprets, structures, and translates intricate tax regulations into executable code. The iterative feedback loops within the agentic framework, particularly driven by the Metamorphic Agent, enable continuous refinement and ensure high reliability and accuracy, especially in handling deductions, credits, and retirement distributions. 
Future work can extend our agentic approach to other legal-critical software, such as poverty management systems, offering a robust pathway for translating complex legal texts into dependable executable software.

At the same time, our study highlights important challenges. Test case generation in legally critical domains continues to suffer from the oracle problem, where correct outputs are difficult to determine without expert interpretation. While higher-order metamorphic testing mitigates this, it also introduces significant computational overhead, as shown in our token usage analysis. Moreover, although smaller models like \texttt{GPT-4o-mini} can outperform frontier LLMs under our agentic framework, scaling these methods to other legal domains will require careful balance between accuracy, efficiency, and interpretability. 

Ultimately, this research points toward a broader vision: LLM-driven agentic methodologies that combine domain expertise with systematic testing can translate complex legal rules into transparent, verifiable, and trustworthy software, advancing accountability and accessibility in critical public services.

\begin{acks}
    This project has been partially supported by NSF under grants CCF-$2532965$, CCF-$2317206$, and CCF-$2317207$.
    Ashutosh Trivedi is a Royal Society Wolfson Visiting Fellow and gratefully acknowledges the support of the Wolfson Foundation and the Royal Society.
\end{acks}

\newpage

\bibliographystyle{ACM-Reference-Format}
\bibliography{references}

\end{document}